\input amstex.tex
\documentstyle{amsppt}

\advance\vsize -2cm
\magnification=1200
\hfuzz=30pt
\NoBlackBoxes

\topmatter
\title The Universal Gerbe, Dixmier-Douady class, and gauge theory \endtitle
\author Alan L. Carey and Jouko Mickelsson \endauthor

\affil Department of Mathematics, University of Adelaide,
SA 5005, e-mail acarey{\@}maths.adelaide.edu.au; 
Mathematical Physics, 
Royal Institute of Technology, SCFAB, 
{SE-106 91} , Stockholm, Sweden, e-mail jouko{\@}theophys.kth.se \endaffil

\date July 23, 2001 \enddate
\endtopmatter

\document

\define\tr{\text{tr}}

\noindent ABSTRACT. We clarify the relation between the Dixmier-Douady class on
the space of self adjoint Fredholm operators (`universal B-field')
and the curvature of
determinant bundles over infinite-dimensional Grassmannians. In
particular, in the case of Dirac type operators on a three dimensional
compact manifold we obtain a simple and explicit expression for both
forms.

\vskip 0.3in
\noindent 0. INTRODUCTION.  \newline\newline

Gerbes arise naturally in quantum field theory (QFT) in several different ways
[Br] [CMM1]. The most recent manifestation is in string theory where it has
been realised that the so-called $B$-field (a family of local 2-forms
$\{B_i\}$
each defined on an open set $U_i$ in a cover of a space-time manifold $M$)
actually defines the Dixmier-Douady (DD) class of a gerbe [FW], [K], [BM],
[HM]. The DD class
is an element of $H^3(M, {\Bbb Z})$ (Cech cohomology).
When this class is not torsion it may also be given by a closed de Rham form
${H_3}$ defined locally on $U_i$ by
$dB_i ={H_3} \vert_{U_i}$. In this letter we will be concerned only with
the non-torsion case.

Now it is well known that
classes in $H^3(M, \Bbb Z)$ classify principal bundles over $M$ with fibre
the projective unitary group of a complex Hilbert space $H$. Let $P$ be such
a principal bundle over $M$ with DD class represented by a de Rham form $[H_3]$.
In this situation
$[H_3]$ has the interpretation of an obstruction to the existence of a
prolongation of $P$ to a bundle with fibre the group of unitaries, $U(H)$,
on $H$. This situation has recently been discussed in a string theory context,
see e.g. [HM],  where it was suggested that the B-field,
can be obtained from a formula
involving a connection on the bundle $P$ and its curvature form. However, it
is difficult to make the proposed formula mathematically rigorous
since it involves taking traces of manifestly non-traceclass
operators.

Similar considerations have arisen in other contexts in standard quantum
field theory constructions when chiral fermions are coupled to external
fields [Se], [CMM], [CM] and these provide some of the ideas for the present
paper.

In the present article we suggest how the construction in [HM]
(which involves a path integral) of a connection
on $P$ and its curvature should be
replaced by a trace of a characteristic class constructed from a
Bismut-Freed type superconnection. In addition, we point out an
interesting relation between the 3-form DD class and the curvature
form of the basic complex line bundle over Grassmannians modelled by
$L_{p+}$ Schatten ideals.  (This latter line bundle is
familiar from the study of anomalies in gauge QFT where the curvature gives
the so-called commutator anomaly of the gauge symmetry group. It is discussed
in [CMM], [MR])

We view the DD class on a given space-time manifold $M$ as being
pulled back from a universal DD class initially
defined on the space of self adjoint
Fredholm operators. Our main result is to give an explicit formula for this
universal class using a Bismut-Freed superconnection.  However to obtain this
formula we need to exploit an observation of Atiyah and Singer and replace
the space of self adjoint Fredholm operators by the unitary operators which
differ from the identity by a compact operator. In fact, we further replace
compact operators by $L_{p+}$ operators in order to be able to write explicit
formulae for de Rham forms as traces of powers of a curvature form taking
values in a Lie algebra of linear operators in a Hilbert space.

The basic problem with the proposal in [HM] is that the group $PU(H)$ is too
large for a geometric analysis of this problem. Actually, QFT gives more
structure.  Elements of $PU(H),$
interpreted as Bogoliubov automorphisms of the fermion algebra,
cannot in general be lifted to a fermionic Fock space. Instead, it is
known that an operator $g$ on $H$ can be lifted to the fermionic Fock space
over
$H$ if and only if it belongs to the \it restricted unitary group \rm on $H$,
denoted $U_{res}$
(for details see [PS]).  The restricted unitaries
act as projective unitaries in the Fock space suggesting that $U_{res}$
may provide a substitute for $PU(H)$. Note that the
connection with anomalies in gauge theories is via the fact that the group
2-cocycle on $U_{res}$ given by this projective action
is infinitesimally just the commutator anomaly in the Lie algebra
(Schwinger term).

The discussion starts in Section 1 with a suitable choice of the space
carrying
the universal Dixmier Douady class (this space is homotopy equivalent to
the set of self adjoint Fredholm operators which are neither
essentially positive
 nor essentially negative). Our choice is dictated by the existence for this
 space of a
simple expression for the generator of degree three de Rham cohomology.
Our earlier work [CM] provides a way to connect with quantum field theory
in a geometric fashion bringing in the group $U_{res}$ and noting that
the universal DD class is the DD class of the so-called `lifting bundle gerbe'
for $U_{res}$.

In Section 2 we define
a Bismut-Freed type superconnection whose curvature gives a formula for
this universal DD class. The Bismut-Freed formula gives the de Rham
form of the DD class on the base of a fiber bundle through an
integration over fibers of a Chern class in the total space (which in
turn can be written as a Dixmier trace of an appropriate operator
form). This should be viewed as a replacement for the above mentioned
ill-defined path integral formula.
   A derivation of the Dixmier-Douady class from the Bismut-Freed form of
the families index theorem was discussed recently in [Lo]; in the
case of Dirac operators coupled to vector potentials or metrics his
results agree with the earlier computations in [CMM], [EM].
Since in the present paper we are studying
a special family of Dirac type operators we obtain more explicit formulas.

In Section 3 we specialise to Dirac operators
on odd dimensional manifolds (and 3-manifolds in particular) where we
can calculate more  explicit formulae for this DD class using suitable
renormalised traces.

\newpage
\noindent 1. A UNIVERSAL GAUGE GROUP. \newline\newline

Let $\Cal F_*$ be the set of all self-adjoint Fredholm operators in an
infinite-dimensional complex
Hilbert space $H,$ with both positive and negative essential
spectrum. According to [AS] the space $\Cal F_*$ is homotopy equivalent
to the set of unitaries $g$ of the type $-1 +$ a compact operator.
The homotopy equivalence is constructed in several steps. The effect of
the first stage is that $\Cal F_*$ is a retract of self-adjoint operators
with essential spectrum at $\pm 1.$ (Actually, Atiyah and Singer talk about
skew self-adjoint operators with essential spectrum $\pm i.$)
In the last step such an operator $F$ is mapped to the unitary operator
$g=\exp(i\pi F)$ with essential spectrum at $-1.$ We prefer to use the
parametrization $g=-\exp(i\pi F)$ and thus $g$ belongs to the group $G_c$
consisting of unitaries such that $g-1$ is compact.

In this paper we shall restrict to the subspace of \it Dirac type \rm
Fredholm operators $\Cal F_D$ and we can define the homotopy equivalence
in a slightly more direct way.
An unbounded self-adjoint operator $D$ is Dirac type if $|D|^{-p}$
is trace-class for some $p\geq 1.$ The infimum of such $p$ is called the dimension
of the underlying (noncommutative) space, [Co].
Let $D_0 \in \Cal F_D$ be fixed and $D_A= D_0 +A$ a bounded perturbation of
$D_0.$ A generalized gauge transformation is a unitary map $g:H\to H$
such that $[D_0,g]$ is bounded. Then $g^{-1} D_A g= D_0 +A'$ is another bounded
perturbation with the gauge transformed potential $A' =g^{-1}Ag
+g^{-1}[D_0,g].$

We can modify the construction of [AS], replacing $G_c$ by the group
$G$ of unitaries differing from the identity by a trace class operator. (Note that Palais showed that
$G$ and $G_c$ are homotopy equivalent, see [Q].)
In this modification we have now a map $A\mapsto g_A \in G$ given by
$$A\mapsto D_A \mapsto F_A= D_A/(|D_A|+ \beta(|D_A|)) \mapsto g_A=
-\exp(i\pi F_A),$$
where $\beta$ is any positive smooth function with
$\beta(0)=1$ and $\beta(x)\to 0$ faster than $|x|^{-p}$  as $x\to \infty.$
The map from $\Cal F_*$ to $G$ is essentially the same as the
restriction of the homotopy equivalence in [AS] to Dirac type operators.

The drawback with the composite map described above is that it is
not gauge covariant; it does not define a map from the space of gauge
orbits $\Cal A/\Cal G$ to $G,$ not even in the case when the space of
perturbations $A$ is the space of smooth vector potentials $\Cal A$
and $\Cal G$ is the group of (based) gauge transformations.
Next we shall explain a
modification which gives a gauge covariant map.

Let us specialize to the case of smooth vector potentials $A=A^i
\gamma_i$ on a connected compact spin manifold $M$ and smooth based
gauge tranformations $g\in\Cal G.$ Here the $\gamma_i'$s are the
Dirac gamma matrices with respect to a fixed Riemannian metric on $M,$
$1\leq i\leq \text{dim}M.$
We have $g_{A'} = g^{-1} g_A g$ for any gauge transformation $g$ and
therefore we cannot use $g_A$ to define a map $\Cal A/\Cal G \to G.$
Let $X=\Cal A/\Cal
G,$ the moduli space of gauge connections; this has the structure of a
(Frechet) manifold because $\Cal G$ acts freely on $\Cal A.$ We can
view $\Cal A$ as the total space of a principal $\Cal G$ bundle over
$X.$ We can define an associated principal $U(H)$ bundle $Z$ over $X$
with total space $\Cal A \times_{\Cal G} U(H),$ where $\Cal G$ acts
naturally from the right on gauge potentials and from the left on
$U(H)$ through the embedding $\Cal G\subset U(H)$ via the gauge group
action on square-integrable fermion fields.

Since $U(H)$ is contractible we can define a global section $X\to Z,$
as a map $r: \Cal A\to U(H)$ such that $r(A^g) = g^{-1} r(A)$ and $r(0)=1.$
Set next $\tilde g_A = r(A)^{-1} g_A r(A).$ This map is gauge
invariant, giving a map $\tilde g: X \to G.$

\noindent 1.1 \bf A universal gerbe. \rm             \define\hU{\hat U_{res}}

In [CM] we constructed a universal $U_{res}$ bundle over the base $G.$
The group $U_{res}=U_{res}(H,\epsilon)$ is defined as the group of unitaries
in a Hilbert space $H$ such that the off-diagonal blocks with respect to
a polarization $\epsilon$ are Hilbert-Schmidt.
The construction was motivated by quantum field theoretic considerations.
In particular, a family $X$ of self-adjoint Dirac operators on a manifold
$M$ can be considered as a subset $X\subset G$ through the homotopy
equivalence described above. To each point $A\in X$ one wants to associate
a fermionic Fock space $V_A$ such that the representation of the canonical
anticommutation relations algebra (CAR) in $V_A$ is equivalent to a representation
defined by the `Dirac sea' construction, that is, by the polarization of the
1-particle space $H$ to positive and negative energies with respect to
the Dirac hamiltonian $D_A.$ Since the group $U_{res}$ acts only projectively
in a fixed Fock space, through a central extension
$$1\to S^1 \to \hU \to U_{res} \to 1,$$
the problem of constructing Fock spaces parametrized
by elements of $X$ reduces to the problem of prolonging a principal
$U_{res}$ bundle over $X$ to a principal $\hU$ bundle.  The potential
obstruction to this is the cohomology class on $X$ obtained as the
pull-back of a basic 3-form on $G,$ with respect to the embedding $X\to G.$

The basic 3-form $c_3$ on $G$ is equal to the Wess-Zumino-Witten-Novikov
(WZWN) 3-form
$$c_3 = \frac{1}{24\pi^2} \tr\, (dg g^{-1})^3.\tag1$$
The form is normalized such that it gives the generator of $H^3(G,\Bbb Z).$
The form $c_3$ is the obstruction to prolonging the universal $U_{res}$
bundle over $G$ to a $\hat U_{res}$ bundle. For this reason it is the
\it Dixmier-Douady class of the universal gerbe. \rm

Thus we may pull back
the three form $c_3$ on $G$ to give a three form on $\Cal A/\Cal
G.$ The cohomology class of the pull-back does not depend on the
choice of $r.$  To see this consider a three dimensional cycle $\Sigma
\subset X$ as a 3-disk $D^3\subset \Cal A$ such that the points on the
boundary are gauge related. Let $h_t(A)$ be the deformation of $\tilde
g: D^3\to G$ given by $h_t(A) = r(tA)^{-1}g_A(t) r(tA)^{-1},$ where
$g_A(t)$ is defined by
$$F_A(t)= (D_A+(t-1)\lambda)/( |D_A+(t-1)\lambda | +t\beta(|D_A|) )\tag2$$
and $\lambda$ is any real number not in the spectrum of $D_A$
for $A$ in the
boundary of $D^3$ (this is a well defined
condition as the $D_A$ for $A$ in the boundary
 are all gauge equivalent). Then $h_1(A)=\tilde g_A$ and $h_0(A)= -\exp(i\pi
(D_A-\lambda)/|D_A- \lambda|).$ The integral of $h_0^* c_3$ does not
depend anymore on $r.$

The explicit formula for $\tilde g^* c_3$ does depend on $r,$ and
is complicated to evaluate.  However, it is possible to build a
`universal model' with concrete formulas for the basic differential
forms. The rest of this note is devoted to calculating these formulae.

 In gauge theory based on the group $U(N)$ the group of
gauge transformations $Map_0(M, U(N))$ on an odd dimensional sphere
$M=S^k$ has the homotopy type
of $\Omega G$ in the limit $N\to\infty.$ Here $\Omega G$ is the group of
based smooth loops $f:S^1\to G,$ $f(1)$ is the neutral element in $G.$
Let us replace the gauge
group $\Cal G$ by $\Omega G$ and the contractible space $\Cal A$ by
another contractible space $PG,$ the space of smooth maps $f:[0,2\pi]\to G$
such that $f(0)=1$ and $f^{-1}df$ is periodic at the end
points. Each such $f$ defines a vector potential
$A=f^{-1}df$ on the circle $S^1.$ Conversely, each periodic
vector potential $A$ defines $f$ uniquely by the given initial condition and
the ordinary differential equation $df=f\cdot A.$ The projection
$\pi:PG\to G$ is the
evaluation $f\mapsto f(2\pi).$ The fiber is $\Omega G.$  The pull-back
$\pi^* c_3$ can be written as $\pi^* c_3 = d\theta$ with
$$\theta_f(X,Y) = \int_0^{2\pi}c_3(f^{-1}df,X,Y)= \frac{1}{8\pi^2}
\int_0^{2\pi} \tr\, (f^{-1}df) [X(t),Y(t)].\tag3$$

The group $\Omega G$ can be embedded in the restricted unitary group
$U_{res}=U_{res}(\Cal H, \epsilon)$ in the Hilbert space $\Cal
H=L^2(S^1, H),$ [PS][CM].
Here the polarization $\epsilon$ is the sign of the (discrete) momentum
on the circle (see [PS] for this construction).
The embedding is defined by a point-wise action on $H$-valued
functions and is a homotopy equivalence (as shown in [CM]). The group $U_{res}$ is also
homotopy equivalent to the larger group $U_{cpt}$ defined by the condition
$[\epsilon,g]$ is compact. The fact that it can be considered as the large
$N$ limit of the group of
based $U(N)$ valued gauge transformations over an odd dimensional sphere
is an important motivation in the \lq universal gauge theory' of [Ra].

As shown in [CMM] the form $\theta$ defines an extension of the Lie
algebra of infinitesimal gauge transformations (by the so called
Schwinger terms).  The modified commutator is defined as
$$[X,Y]' = [X,Y] + 2\pi i\theta_f(X,Y),\tag4$$
that is, as a sum of the pointwise commutator of the loop algebra
elements $X,Y$ and a scalar function of $f.$ Thus the extension is
defined by the abelian ideal of functions of $f.$ The cocycle
$\theta$ is equal, modulo a coboundary, to the cocycle
$$c(X,Y)= \frac{i}{2\pi} \int \tr\, X dY.\tag5$$
$c$ is the cocycle defining an affine Lie algebra. The coboundary
relating
the two cocycles is $\delta w$ with $w_f(X)= \frac{i}{4\pi} \int \tr\, A X.$

There is a family of local closed 2-forms on $\Cal A$ which gives
an alternative description of the gerbe arising from the DD class
$c_3.$ First, near the unit element in $G$ we can write
$c_3=d\psi$ with the parametrization $g=-e^{i\pi F},$
$$\psi= \frac{1}{8} \tr\, dF h(ad_{i\pi F}) dF,\tag6$$
where $h(x)=\frac{\sinh(x) -x}{x^2}$ and $ad_X(Y)=[X,Y].$
The local form $\eta= \theta-\pi^*\psi$ is closed since $d$ and the
pull-back
operation commute. Along gauge orbits the forms $\eta$ and $\theta$
are the same.  However, one must remember that $\psi$ is defined only
in the open set where the exponential function is bijective whereas
$\theta$ is globally well-defined on $\Cal A.$
We can cover $\Cal A$ with open sets $V_{\alpha}=\pi^{-1}(U_{\alpha}),
$ where each $U_{\alpha}$
is of the type $U_{\alpha}= g_{\alpha} U_1$ and $U_1$ is an open
neighborhood of the identity in $G.$ Each $U_{\alpha}$ can be parametrized
as $g_{\alpha} e^{i\pi F}$ for $F\in Lie(G).$
In this parametrization the local form $\psi_{\alpha}$ on $G$ has the same
expression as before, near the identity. The curvature of the
local line bundles $L_{\alpha\beta}$ on the overlaps $V_{\alpha\beta}$
is given as $\frac{1}{2\pi}\omega_{\alpha\beta}=
\pi^*(\psi_{\alpha}) -\pi^*(\psi_{\beta})
$ and is closed, again because $d$ and the pull-back operation commute.
Note however that although $\omega_{\alpha\beta}$ can be pushed forward
to $G,$ it is not a difference of closed forms on $G.$ On $\Cal A$ we
have $\omega_{\alpha\beta}= \omega_{\alpha}-\omega_{\beta}$ as a difference
of closed forms with $\frac{1}{2\pi} \omega_{\alpha}= \theta -
\pi^*(\psi_{\alpha}).$
This reflects the fact that on $\Cal A$ the gerbe is necessarily trivial
(but not down on $G$)
which means that  $L_{\alpha\beta}= L_{\alpha}\otimes L_{\beta}^*$
for a family of local line bundles $L_{\alpha}$ over the open sets
$V_{\alpha}.$

\vskip 0.3in
\noindent 2. CONNECTION IN A UNIVERSAL BUNDLE AND THE DD CLASS.\newline\newline

In this section we use the superconnection
formalism of [BF] to give a formula for the universal DD class.
Let $G\subset U(H)$ as before and  let $\pi:P\to G$ be a  principal bundle
over $G$ with
structure group $\Omega G$
and total space $P$ the space of smooth paths in $G$ starting
from the unit element and such that $f^{-1}df$  is periodic  at the end points
$x=0,2\pi$ of the paths $f(x).$ The projection map $\pi$
is $f\mapsto f(2\pi).$
The action of $\Omega G$ on $P$ is the pointwise right multiplication,
$(f\cdot g)(x)= f(x)g(x).$

As before, each element $f\in P$ can be viewed as a smooth vector potential on
the circle $S^1$ (which is parametrized as $e^{ix}$)
of the form $A= f^{-1}df.$

We define a connection on $P.$  For this purpose choose any smooth
real valued function $\alpha$ on the interval $[0,2\pi]$ such that
$\alpha(0)=0,$ $\alpha(2\pi)=1,$ and all the derivatives of $\alpha$
vanish at the end points.  The connection is defined as a $Lie(\Omega
G)$ valued 1-form $\omega$ on $P,$
$$\omega = f^{-1}\delta f -\alpha(x) f(x)^{-1} \left( \delta f(2\pi)
f(2\pi)^{-1} \right) f(x),\tag7$$
where the exterior differentiation on $P$ is denoted by $\delta$ in
contrast to the differentiation $d$ on the circle.

One immediately checks that $\omega$ is indeed a connection form:
along vertical directions defined by the right $\Omega G$ action it
is tautological, $\omega(X)= X$ for $X\in \Omega \bold{g},$ and it is
equivariant,  $r_g\omega= g^{-1} \omega g$ for any $g\in \Omega G.$

Let $Q$ be the $G$ bundle over $S^1\times G$ defined as follows.
Start from $\tilde Q=S^1\times P\times G$ viewed as a trivial $G$ bundle over
$S^1\times P.$ The group $\Omega G$ acts freely from the right as
$$(x,f,a)\cdot g= (x, fg, g(x)^{-1}a).\tag8$$
Thus we may pass to the quotient $Q=\tilde Q/\Omega G$ and this is a
(nontrivial) $G$ bundle over $S^1\times G$ since the right actions of $G$ and
$\Omega G$ commute.

We can extend the connection $\omega$ to a connection in $Q.$ It is a
sum of two terms, $\omega=\omega^{(1,0)} +\omega^{(0,1)},$ where the
first term is a 1-form along $S^1$ given as $f^{-1}df$ in terms of a
local section $G\to P$ which sends $a\in G$ to $f=f_a\in P$ with
$f_a(2\pi)=a.$ The second term  is given by the formula (7) above,
with respect to the same local section.

The curvature is easily computed to be $\Cal F=\Cal F^{(2,0)} +\Cal F^{(1,1)}+
\Cal F^{(0,2)}$ with $\Cal F^{(2,0)}=0$ (since dim$S^1=1$), and
$$\align \Cal F^{(1,1)} &= -d\alpha f^{-1}\left(\delta f(2\pi) f(2\pi)^{-1}\right)
 f \\ \Cal F^{(0,2)} &= (\alpha^2-\alpha) f^{-1} \left(\delta f(2\pi)
f(2\pi)^{-1}\right)^2 f. \tag9 \endalign$$

Following [BF] the odd Chern classes on the base $G$  of the bundle $P$ are
then obtained by integrating along the fiber $S^1$ the ordinary even Chern
classes in the base $S^1\times G$ of the bundle $Q.$ For example, the
3-form part is given as
$$\align c_3&=  \frac{1}{8\pi^2} \int_{S^1} \tr\, (\Cal F^2)^{(1,3)}
= -\frac{1}{4\pi^2} \int_{S^1} \tr\, (d\alpha)(\alpha^2-\alpha)
f(x)^{-1}\left(\delta f(2\pi) f(2\pi)^{-1}\right)^3 f(x) \\
&= \frac{1}{24\pi^2}\tr\, \left(\delta f(2\pi)f(2\pi)^{-1}\right)^3
\tag10\endalign$$
which is the canonical 3-form $\frac{1}{24\pi^2}
\tr(da a ^{-1})^3$ on the group $G.$ The higher forms
$$\tr\,(da a^{-1})^{2k+1}$$
 are obtained in exactly the
same way, by starting from the higher Chern classes $\tr \Cal F^{k+1}.$

The (1,1) form part of the curvature can be also written as
$$\Cal F^{(1,1)} =  [D_f, \delta+ \omega] $$
where $D_f$ is the (antihermitean) Dirac operator on $S^1,$ $D_f =
\frac{d}{dx} + A_f,$ with $A_f =f(x)^{-1}f'(x).$
The three form can be written as
$$c_3  = \frac{1}{8\pi^2}\cdot  2\pi\tr_+ \,\frac{1}{|D|}
(\Cal F^2)^{(1,3)}.\tag11$$
Here $\tr_+$ is the Dixmier trace.  Note that all components of the
curvature are bounded operators in the space $L^2(S^1, H).$

This way of writing the 3-form leads immediately to a generalization.
Let $p$ be an odd positive integer.
Assume that $|D|^{-p}$ is in the \lq Dixmier ideal'
$L_{1+},$ with $D$ a linear operator in a
dense domain in a Hilbert space $\Cal H.$ (In the previous example
$\Cal H= L^2(S^1,H)$) We assume that we have a bigraded differential
algebra $\Omega^{(*,*)}$ with a differential $\Delta= d + \delta,$
where $\delta$ is the de Rham differential in the base space as
before, increasing the second index by one unit, and $d$ is a
generalized differentiation in the sense of noncommutative geometry (NCG),
anticommuting with $\delta.$ The $k-$forms in the $d$ complex are
constructed as linear combinations of operators of the form $\phi=
a_0 da_1\dots da_k,$ where $da= [D,a]$ and the $a_i$'s are elements in an
associative algebra of bounded operators such that the commutators
$[D,a_i]$ are bounded, [Co]. Naively, the exterior derivative of $\phi$
is equal to $da_0 da_1\dots da_k.$ However, there is an ambiguity in
the expression for $\phi,$ as a differential polynomial in the algebra
elements $a_i,$ and therefore one has to pass to an appropriate
quotient algebra; see [Co], Section VI.1,  for details.

Next we assume that  $\omega=\omega^{(1,0)} +\omega^{(0,1)}$ is a
graded 1-form on the base of the type described above and we denote by
$\Cal F$ its graded curvature form.
For any positive odd $k$ we have a closed
form on the base defined by
$$ c_k = \frac{1}{n! (2\pi)^n} b_p\tr_+ \,\frac{1}{|D|^p}
(\Cal F^n)^{(p,k)},\tag12$$
where $2n=p+k$ and $b_p=  (2\pi)^p/\text{vol}(S^{p-1})$ is a
normalization
constant; the normalization is chosen such that $b_p \tr_+ \frac{h}{|D|^p}=
\int_M h(x) dx$ when $h$ is a multiplication operator by a smooth
function in the spin bundle over the compact manifold $M$ of dimension
$p=2m+1.$

\noindent 2.1 \bf A Basic Example. \rm  Replace $G$ above by $\Cal G=Map_0(S^{2m},G),$
the group of based maps. For a topologist, $\Cal G$ is the same as $G$
but for the following geometric analysis there is a difference. Set
$P=Map_0(B^{2m+1},G),$ where $B^{2m+1}$ is a disk with boundary
$S^{2m}.$ The maps are based on a radius connecting the origin of the
disk to the base point on $S^{2m}.$ We can view $P$ as a principal
bundle over $\Cal G,$ the projection is the restriction onto the
boundary and the fiber is equal to the group of based maps from
$S^{2m+1}$ to $G;$ we identify functions on the disk $B^{2n+1}$ which are
constant on the boundary as functions on $S^{2n+1}.$
This bundle has a connection defined as above,
$$\omega= f^{-1}\delta f -\alpha(r) f^{-1} \left( \delta f(r=1)
f(1)^{-1} \right)f,\tag13$$
where $\alpha$ is a smooth function of the radius $r$ alone and
otherwise it satisfies the same conditions as in the 1-dimensional
case. There is a superconnection defined as $A=A^{(1,0)} + A^{(0,1)}$
where $A^{(0,1)}=\omega$ and $A^{(1,0)}$ is a $Lie(G)$ valued vector
potential on $S^{2m+1}$ defined by
 $$  A^{(1,0)} = f^{-1} df -\alpha(r)f^{-1}\left( df(1)
f(1)^{-1})\right) f.\tag14$$
Here $d$ is just the classical de Rham differentiation. It is a
special case of the NCG differential operator $d;$ for example, a
vector potential $A^i dx_i$ corresponds to the operator $A^i\gamma_i, $
which can be written as a linear combination of operators $a_0[D,a_1]$
where $a_0,a_1$ are multplication operators by smooth functions and
$D$ is the Dirac operator determined by the standard metric on the
disk $B^{2m+1}.$

Thus the superconnection can be written as
$$A= f^{-1} \Delta f -\alpha(r) f^{-1} \left( \Delta f(1)
f(1)^{-1}\right) f,\tag15$$
where we denote $\Delta=d +\delta.$
The curvature of this superconnection is
$$ \Cal F= -d\alpha f^{-1}\left( \Delta f(1) f(1)^{-1}\right) f
+(\alpha^2-\alpha) f^{-1}\left(\Delta f(1) f(1)^{-1}\right)^2 f.\tag16$$
It can be decomposed as $\Cal F=\Cal F^{(2,0)} +\Cal F^{(1,1)} +\Cal F^{(0,2)}.$
The Chern character is generated by the powers
$\tr \Cal F^k$
and the forms on the base $\Cal G$ of the bundle $P$ are given as
$$ c_k = \frac{1}{n!(2\pi)^n} \int_{B^{2m+1}} \tr\, (\Cal F^n)^{(2m+1,k)},\tag17$$
with $2n= 2m+1+k.$ In fact, by Stokes' theorem the integral reduces
to a boundary integral over $S^{2m},$ since the integrand is of the
form
$$d\alpha (\alpha^2-\alpha)^{n-1} \tr\, (\Delta f(1)
f(1)^{-1})^{2n-1}$$
and the radial integration gives just the numerical
factor
$$\int_0^1 \alpha'(r) (\alpha^2-\alpha)^{n-1} dr.$$
Thus
$$c_k = N_{k,m} \int_{S^{2m}} \tr\, ((\Delta f(1) f(1)^{-1})^{2n-1})^{(2m,k)},
\tag18$$
where $N_{k,m}$ is a normalization constant.
In the case $m=0$ the integration reduces to a calculation of $f(1)$
in a single point and we recover the formula in the case of the
universal bundle over $G.$

We observe that the integration in eq. (17) can be replaced, up to a
normalization constant,  by the Dixmier
trace $\tr_+ \frac{1}{|D|^p}(\Cal F^n)^{(2m+1,k)}$ with $p=2m+1.$ Thus
eq. (17) is indeed a special case of (12).

\newpage
\noindent 3. AN APPLICATION TO GAUGE GROUPS
 AND GERBES IN ODD DIMENSIONS.
   \newline\newline

Let $\Cal A$ be the space of $\bold g$ valued vector potentials on a
compact spin manifold $M$ of dimension $p=2k+1.$ Let $D_0$ be a fixed Dirac
operator (defined, for example, with the vector potential
$A=0$). Denote $D_A= D_0 +A$ with $A= i\sum \gamma^k A_k$ for $A\in \Cal
A.$ Denote $F_A= D_A (D_A^2 +m^2)^{-1/2}$ where $m$ is a positive constant.
Then $F_A-\epsilon \in
L_{p+},$ where $\epsilon= D_0/|D_0|;$ if zero is an eigenvalue of $D_0$ we put it on the positive side
in the spectrum. To each $D_A$ we can associate
the unitary operator $g_A = -\exp{i\pi F_A}.$
We claim that this operator differs
from the  unit operator by a perturbation which is in the ideal  $L_{p+}$
consisting of operators $T$ with $|T|^p \in L_{1+}.$
To see this we introduce
$\epsilon_A= D_A/|D_A|$, with the same convention for zero eigenvalues as
for $\epsilon$, and note that $g_A=\exp{i\pi \epsilon_A(|F_A|-1)}$
so that it suffices to prove that $|F_A|-1\in L_{p+}$.
Using $|F_A|-1=(1+|F_A|)^{-1}(F_A^2-1)$ and $F_A^2-1= -m^2\cdot
(D_A^2 +m^2)^{-1}$
we have
$$(D_A^2 +m^2)^{-1} = D_A^{-2} - m^2(D_A^2 +m^2)^{-1}D_A^{-2}$$
and the RHS clearly lies in $L_{p+}$ so that  $|F_A|-1$ lies in $L_{p+}$.

Thus in this setting we need to extend the forms $c_j$ in the previous section to the
the group $U^{(p+)}$ of unitaries $g$ such that $g-1\in L_{p+}.$
(Note that if $U^{(p+)}$ is
equipped with the Banach Lie group structure arising from the norm
on $L_{p+}$ it is not homotopy equivalent to $G$
as the former is not separable. This will not be an issue in
our discussion.)
For that purpose we use the method explained in [LMR].
Replace $g$ by a $2\times 2$ operator matrix and define the
operators $\epsilon,\Gamma,$
$$g\mapsto \tilde g=\left(\matrix g & 0\\ 0& 1\endmatrix\right) \hskip
0.3in
\sigma = \left(\matrix 0 &1\\ -1&0\endmatrix\right) \hskip 0.3in
\Gamma=\left(\matrix 1&0\\0& -1\endmatrix\right).$$
Replace the variation $\delta$ by $\pm \Gamma\delta$ and denote
$d\tilde g=[\sigma,\tilde g].$ We have once again a NCG bigraded
differential algebra and a total differential $\Delta=d +\delta,$
where now the first differential $d$ is defined as $d \tilde g =
[\sigma, \tilde g]$ for 0-forms, and in general
$d\phi= d\alpha_0 d\alpha_1\dots d\alpha_j$ for $\phi=a_0 da_1 \dots
da_j$ for j-forms $\phi.$ Note that in the present setting
$d\phi=[\sigma, \phi]$ for even degree forms and
$d\phi=\sigma\phi+\phi\sigma$ for odd  forms.  One must select the
sign convention in the definition of $\delta$ on odd/even forms such
that $d,\delta$ anticommute, giving $\Delta^2=0.$

For any odd positive integers $j$ and $q=2l+1>p$ we can now set
$$c_{j,q} = -(\frac{1}{2\pi i})^{l+1}\frac{l!}{(2l+1)!} \tr \,\Gamma
\left((\tilde g^{-1}\Delta \tilde g)^{q} \right)^{(q-j,j)}\tag19$$
and this is a closed degree $j$-form on $U^{(p+)}.$ For example, for $j=1$
this expression is, up to a constant, equal to
$$\tr (g+g^{-1}-2)^{(q-1)/2} g^{-1}\delta g.$$
In the case of even $q$ the trace vanishes identically. In the case
of $g-1\in L_1$ we can set $q=1$ and we get the standard expression
$\frac{-1}{2\pi i} \tr(g^{-1}\delta g)$ for the
1-form on $G.$ When $p=j$ we have $c_j \sim \tr (g^{-1}\delta g)^j,$
which is logarithmically diverging but can be replaced by a
renormalized trace, see below.

The case of $q>j >1$ gives a slightly more complicated expression. The
reason is that the operators $\tilde g^{-1}\delta \tilde g$ and
$\tilde g^{-1} d\tilde g$ do not commute and the trace has to be
computed separately for all possible combinations of products of these
operators.  However, for pseudodifferential operators (PSDO's) on a compact
manifold  the borderline case $q=j=2k+1$ still gives a nice
formula,
$$c_{p,p} = -(\frac{1}{2\pi i})^k \frac{k!}{(2k+1)!} \text{TR}
(g^{-1}\delta g)^p,\tag20$$
where the renormalized trace TR is defined as follows. Fix a positive PSDO
$Q$ of order one. Then
$$\zeta(z;T)= \tr\, (Q^{-z} T)\tag21$$
is defined and analytic when Re$(z)$ is large and positive. For a classical
PSDO $T$ the trace at $z=0$ has at most a simple pole; subtracting the pole
one gets a finite expression and this is defined as TR$\,T.$ (In physics
literature this is called the dimensional regularization.)
The following lemma is proven in [CDMP]:

\proclaim{Lemma} If $A,B$ is a pair of PSDO's on a compact manifold $M$ such
that the order of $AB$ is less or equal to $- \text{dim}\, M$
then TR$[A,B] =0.$ \endproclaim

When $g=-\exp(i\pi F_A)$ we can define a 1-parameter family
$$g_t=
-\exp(i\pi(1-t)\epsilon  +i\pi t F_A)\tag22$$
in the group $U^{(p+)}.$ As in the proof
of Poincare's lemma we can define a form $q_{j-1,p}$ such that $dq_{j-1,p}
= c_{j,p}$ by the formula
$$q_{j-1,p} =  \int_0^1 i_{i\pi(F_A- \epsilon)} c_{j,p}(g_t) dt ,\tag23$$
where $i_X$ is the interior product of a vector field $X$ with a
differential form.

In the case of Dirac operators coupled to vector potentials on a compact
manifold $M$ of dimension $3$ we can be a bit more explicit. In this case
$p=3$ and using the above lemma we can set $j=3.$ By (6) we have now
$c_3= c_{3,3}=d\psi;$ note that
in the present case $F_A -\epsilon \in L_{3+}.$ In particular, if
$F^2=1$ the form $\psi$ reduces to
$$\psi = \frac{1}{16\pi} \text{TR}\, F (\delta F)( \delta F), \text{ for }
F^2=1.\tag24$$
This is the curvature form of the determinant line bundle over the
$L_{3+}$ Grassmannian.  It gives the Schwinger terms when evaluated
along gauge directions, [MR], [LM].
Here gauge transformations
are of the form $F\mapsto aFa^{-1},$ with $a$ unitary.
Denoting by $X,Y$ a pair of infinitesimal gauge transformations
the Schwinger term becomes
$$s(F;X,Y)=  \frac{1}{4}  \text{TR}\, F [X,F][Y,F].\tag25$$
Up to a coboundary, this is the same as in [MR],
$$s'(F;X,Y)= \frac18 \text{TR} (F- \epsilon) [[\epsilon,X],[\epsilon,Y]].
\tag26$$

In the case when $F-\epsilon$ is trace-class the difference $s-s'$
is equal to the coboundary of the Lie algebra 1-cochain
$$\eta(X;F)= \frac12 \tr \{ (\epsilon -F)X +\frac14 F \epsilon [\epsilon,X]\}.$$

One can derive the cohomology classes $c_{j,p}$ on the base $U^{(p+)}$ as
in the case of trace class perturbations of the unit operator, Section
2.

We can define a graded 1-form $A$ on the path space $\Cal P$ of  $U^{(p+)}.$
As before, we consider paths $f(x)$ (with $0\leq x\leq 2\pi$)
 starting from $f(0)=1$ and with
periodic boundary conditions on $f^{-1}df.$
The form $A$ is defined as
$$\align A^{(1,0)} & = \tilde f^{-1}{\hat d} \tilde f -\alpha(x)
\tilde f^{-1}  \left( [\sigma, \tilde f(2\pi)]
\tilde f(2\pi)^{-1}\right) \tilde f \\
A^{(0,1)} & = \tilde f^{-1} \delta \tilde f -\alpha(x) \tilde f^{-1}
\left( \delta \tilde f(2\pi) \tilde f(2\pi)^{-1}\right) \tilde f.\endalign $$
Here $\hat d \tilde f(x)  =  \tilde f'(x)dx +[\sigma,\tilde f(x)].$
For higher order $(\sigma,\delta)$ forms one must associate a sign
$s=\pm 1$ with the symbol $dx$ in order to guarantee
that  the various differentials anticommute.
Note that $A^{(1,0)}$ can be further split into a form $\tilde f^{-1}
\tilde f' dx$ along the circle parametrized by
$x,$ and the Fredholm module form
$$\tilde f^{-1}[\sigma,\tilde f] - \tilde f^{-1}\left(
[\sigma, \tilde f(2\pi)] \tilde f(2\pi)^{-1}\right) \tilde f.$$

The  forms $\tr\,\Cal F^n,$ where $n/2>p$ and $\Cal F$ is the
curvature of  the connection $A,$
can be 'integrated' to
give  odd forms on the base. Integration is to be understood in the
NCG sense: It involves a true integration along the circle and an
operator trace in the algebra of 2 by 2 operator matrices; the entries
are operators in the Hilbert space $H$ where the action of $U^{(p+)}$
is defined.  The $x-$ integration is easily performed as in the
section 2 to give the formula
$$c_{j,2n-1} = -(\frac{1}{2\pi i})^{n} \frac{(n-1)!}{(2n-1)!} \text{tr}\,
(\tilde f(2\pi)^{-1} \Delta  \tilde f(2\pi))^{(2n-1-j,j)}\tag27$$
which agrees with (19), with $\tilde g$ replaced by $\tilde f(2\pi)$ and
$\Delta g = \delta g +[\sigma,g]$ and $q=2n-1;$ in the case of PSDO's
on a compact manifold we can go down to the borderline case $q=p$ by
replacing the trace `tr' by the renormalized trace `TR'.

\vskip 0.3in
\noindent \bf Acknowledgement \rm This work was supported in part by the G\"oran
Gustafsson foundation and the Australian Research Council.
The first named author completed this research in part for the Clay Mathematics
Institute.

\vskip 0.3in

\noindent \bf References. \rm \newline\newline

[AS] M.F. Atiyah and I. Singer: Index theory for skew-adjoint Fredholm
operators. I.H.E.S. Publ. Math. \bf 37, \rm 305 (1969)

[BF]  J.-M. Bismut and  D. Freed: The analysis of elliptic families. II.
Dirac operators,  eta invariants, and the holonomy theorem. Commun. Math.
Phys. \bf 107, \rm 103-163 (1986)

[BM]
P. Bouwknegt and V. Mathai,
{\it $D$-branes, $B$-fields and twisted $K$-theory},
J. High Energy Phys. {\bf 03} (2000) 007, hep-th/0002023.

[Br]
J.-L. Brylinski, {\it Loop spaces, characteristic classes and
geometric quantization},
Prog. Math., {\bf 107}, Birkh\"auser, Boston (1993).

[CDMP] A. Cardona, C. Ducourtioux, J.P. Magnot, and S. Paycha: Weighted
traces on algebras of pseudodifferential operators and geometry of loop
groups. math.OA/0001117

[CMM] A.L. Carey, J. Mickelsson and M.K. Murray: Index theory, gerbes, and
Hamiltonian quantization. Commun. Math. Phys. \bf 183, \rm 707 (1997);
hep-th/9511151

[CMM1] A. L. Carey, J. Mickelsson, and M. Murray:
 Bundle gerbes applied to quantum field
theory. Rev. Math. Phys. \bf 12, \rm 65 (2000); hep-th/9711133

[CM] A.L. Carey and J. Mickelsson: A gerbe obstruction to quantization of
fermions on odd-dimensional manifolds with boundary. Lett. Math. Phys.
\bf 51, \rm 145 (2000); hep-th/9912003

[Co] A. Connes: \it Noncommutative Geometry. \rm  Academic Press, London (1994)

[CrMi] C. Cronstr\"om and J. Mickelsson: On topological boundary
characteristics in nonabelian gauge theory.
J. Math. Phys. \bf 24, \rm 2528 (1983)

[EM] C. Ekstrand and J. Mickelsson: Gravitational anomalies, gerbes, and
hamiltonian quantization. hep-th/9904189. Commun. Math. Phys. \bf 212,
\rm 613 (2000)

[FW] D.~Freed and E.~Witten
{\it Anomalies in String Theory with D-Branes}, hep-th/9907189.

[Ha] J. Harvey: Topology of the gauge group in noncommutative gauge theory.
hep-th/0105242

[HM] J. Harvey and G. Moore: Noncommutative tachyons and K-theory. Preprint,
hep-th/0009030

[Kap] A. Kapustin:
$D$-branes in a topologically non-trivial $B$-field.
Adv.\ Theor.\ Math.\ Phys.\ {\bf 4} (2001) 127, hep-th/9909089.

[LM] E. Langmann and J. Mickelsson: $3+1$-dimensional Schwinger terms and
noncommutative geometry. Phys. Lett. \bf B 338, \rm 241 (1994)

[LMR] E. Langmann, J. Mickelsson and S. Rydh: Anomalies and Schwinger terms
in NCG field theory models. J. Math. Phys. \bf 42, \rm 4779 (2001), 
hep-th/0103006.

[Lo] J. Lott: The index gerbe. math.DG/0106177

[MR] J. Mickelsson and S. Rajeev: Current algebras in $d+1$
dimensions and determinant bundles over infinite-dimensional Grassmannians.
Commun. Math. Phys. \bf 116, \rm 365 (1988)

[PS] A. Pressley and G. Segal: \it Loop Groups. \rm Clarendon Press,
Oxford (1986)

[Q] D. Quillen:  Superconnection character forms
and the Cayley transform, Topology \rm {\bf 27} 1988, 211-238 and
P de la Harpe: \it Classical Banach-Lie algebras and Banach Lie groups
of operators in Hilbert Space,
\rm Springer Lecture Notes in Mathematics {\bf 285} 1972

[Ra] S. Rajeev: Universal gauge theory. Phys. Rev. \bf D 42, \rm 2779 (1990)

[Se] G. Segal: Faddeev's anomaly and the Gauss's law. Preprint (1986)

\enddocument